# Effect of Blast Exposure on Gene-Gene Interactions

Akram Yazdani

Department of Neuroscience, Icahn School of Medicine at Mount Sinai, New York, NY, USA

**Abstract**

Repeated exposure to low-level blast may initiate a range of adverse health problem such as traumatic brain injury (TBI). Although many studies successfully identified genes associated with TBI, yet the cellular mechanisms underpinning TBI are not fully elucidated. In this study, we investigated underlying relationship among genes through constructing transcript Bayesian networks using RNA-seq data. The data for pre- and post-blast transcripts, which were collected on 33 individuals in Army training program, combined with our system approach provide unique opportunity to investigate the effect of blast-wave exposure on gene-gene interactions. Digging into the networks, we identified four subnetworks related to immune system and inflammatory process that are disrupted due to the exposure. Among genes with relatively high fold change in their transcript expression level, *ATP6V1G1, B2M, BCL2A1, PELI, S100A8, TRIM58* and *ZNF654* showed major impact on the dysregulation of the gene-gene interactions. This study reveals how repeated exposures to traumatic conditions increase the level of fold change of transcript expression and hypothesizes new targets for further experimental studies.

**Introduction**

Traumatic brain injury (TBI) constitutes a major global health and socio-economic problem. TBI is a disease process with an initial injury that induces biochemical and cellular changes which in turn contribute to continuing neuronal damage and death over time. TBI is a significant cause of death and disease in the armed forces. It has been estimated that 10-20% of veterans have suffered a TBI, and in many cases were not even diagnosed prior to discharge. Psychopathological symptoms associated with TBI include mood and anxiety disorders, post-traumatic stress disorder, suicidality, and diminished cognitive capacity with deficits in attention and memory. These symptoms can arise years after time of injury, leading to misdiagnosis and lack of proper care and treatment.

Emerging evidence reveals that repeated exposure to blast may lead to traumatic brain injury (TBI) symptomatology and related neuropsychiatric sequelae (Elder, Stone, & Ahlers, 2014). In this effort, we have undertaken human studies involving "breachers" military and law enforcement personnel who are exposed to repeated blasts as part of their occupational duty. To achieve the element of surprise and maintain tactical advantage, breachers are typically in close proximity to controlled, low-level blast during explosive breaching operations and training, repeatedly being exposed to blast overpressure waves. Breachers have reported a range of physical, emotional, and cognitive symptoms, including headache, sleep issues, anxiety, and lower cognitive performance (Carr et al, 2015). Although no apparent physical damage presented virtually due to low-level blast exposure, it activates multiple apoptotic and inflammatory pathways through acute effects on the brain (Carr et al., 2015). Furthermore, many studies provide evidence that the effect of exposure to blast is cumulative and long-lasting (Elder et al., 2014).

While TBI is a complex disease and multiple pathways are involved in the pathology and mechanism of the disease, complementary studies to experimental studies are required. Overwhelming studies based on microarrays (Redell et al., 2013; Rojo et al., 2011; Samal et al., 2015) and more recently RNA sequencing (Lipponen, Paananen, Puhakka, & Pitkänen, 2016) focus on differential gene expression in response to TBI and successfully identified biomarkers associated to TBI. However, those studies do not provide information regarding changes in the system while analyzing each gene separately regardless of their relationships. To address this limitation, approaches like network analysis are applied.

Biological networks allow us to uncover the inherent gene-gene interactions, leading to identification of functional relationships. Environmental insults, such as blast, induce changes in the gene network which will allow us to uncover key gene targets that are modulated as a result of exposure to blast. One of the well-stablished approaches to uncover the interaction among genes is Bayesian networks (Pearl, J. 2011) constructed from observational data. Bayesian networks discover the most probable targeting interaction since it is based on conditional dependency. Each connection is established after excluding the effect of other genes in the analysis. These networks also generate new hypotheses for design of follow-up experiments.

In this study, we conducted a Bayesian network to investigate the effect of acute exposure to blast and the transcriptional response to acute blast exposure and how this varies based on lifetime history of TBI. In addition, we aimed to predict essential disruption in the gene-gene interactions due to pressure waves produced by the blast. We constructed the network on RNA-seq data collected from 32 individuals participating in a 2-week cycle at U.S. army explosive entry training sites. The blood samples of the individuals were collected pre- and post-training to profile transcript abundance which makes possible comparison between pre- and post-identified networks. This systems approach combined with specific data set available in our study provided a unique opportunity to study the mechanism of disease through identifying gene-gene interaction altered by blast exposure given reported TBI lifetime history.

**Materials and methods**

**Samples and subjects:** This study includes 32 samples from site 3 of a study designed by Naval Medical Research Center and Walter Reed Army Institute of Research Institutional Review Boards. In this training program, all participates are male, 19 participants reported lifetime history of TBI, and 24 reported prior experience with explosive breaching. Figure 1, right plot, represents the overlap between subsets of individuals with a lifetime history of TBI and a lifetime history of breaching in a Venn diagram. All individuals who were enrolled as students or designated as instructors for the training programs were eligible to participate in the study. The average age of participants is 31 years with a standard deviation 4.6 years, Figure 1, left plot. During the 2-week period, trainees were exposed to multiple blasts and their blood was taken once at baseline and once at the end of training program.

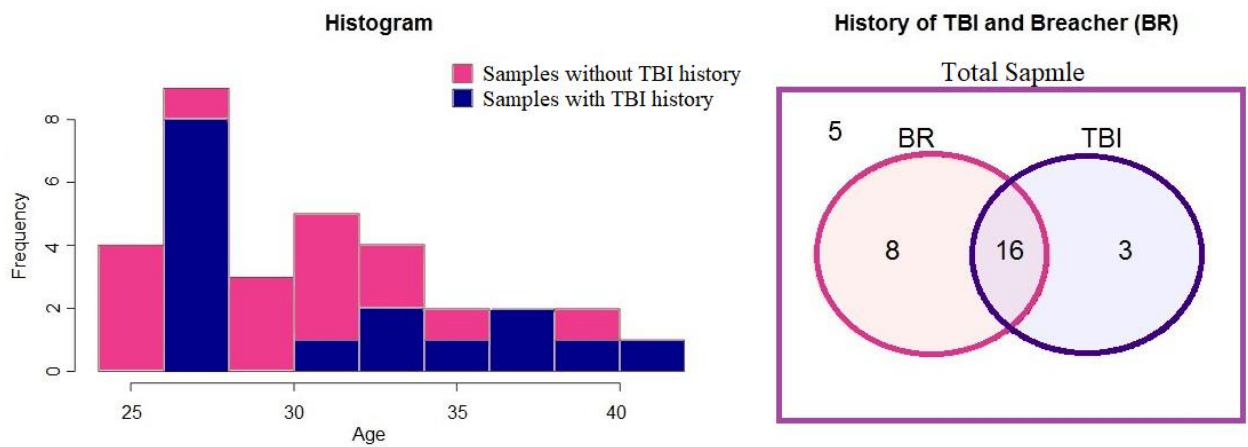

**Figure 1**- Left: Histogram of age. Right: Venn diagram for lifetime history of TBI and breacher (BR).

**RNA processing and sequencing:** Total RNA sequencing libraries were prepared using the Illumina Stranded at baseline (pre; day 1 prior to blast exposure) and at termination of blast training (post; following blast exposure). Final libraries were evaluated using PicoGreen (Life Technologies) and Fragment Analyzer (Advanced Analytics) and were sequenced on an Illumina HiSeq2500 sequencer (v4 chemistry) using 2 x 125bp read length.

**RNA data processing:** RNA-seq reads were aligned to the human reference hg19 using STAR aligner (v2.4.0c) (Dobin et al., 2013). Quantification of genes annotated according to Gencode v18 was performed using feature Counts (v1.4.3). The QC metrics were collected with Picard (v1.83) and RSeQC (Wang, Wang, & Li, 2012).

**Network analysis:** Having pre- and post-blast gene transcript data for the same individuals, we constructed the gene expression Bayesian network (Friedman, Geiger, & Goldszmit, 1997) and identified changes in gene-gene interaction related to blast exposure. The input data for this analysis was the count matrix transcriptome reads that was normalized for gene length and sequencing depth. We selected 2,578 genes from the pre- and post-blast datasets. Only genes with at least 10 samples with nonzero reads and standard deviation greater than 2 were considered. The Bayesian networks is constructed separately for the pre- and post- datasets. Briefly, a Bayesian network is a probabilistic network that determines the conditional dependencies between genes in the system. This is accomplished by calculating joint distribution across all selected genes which can be factorized as probabilities of expression of an individual gene condition on all the selected genes in the analysis. In the RNA-seq networks, each node stands for a gene and the edge between nodes represents interaction between corresponding nodes. To assess quality of the fit in identified networks, we employed hamming distance (Lindell, 2010). The Hamming distance between two networks quantifies the difference between the networks. Here we apply this distance to measure the number of direct connections of each gene that are in common between the two networks. Calculating hamming distance for different statistical significant levels (0.0001, 0.001, 0.01), we found 0.001 as the best significant level to construct the networks in our analysis.

After constructing pre- and post-blast networks, we compared these to identify changes due to blast exposure. We identified subnetworks associated with blast exposure that contained at least 4 genes in the pre-blast network, in which the gene-gene connectivity was disrupted (less than 2 connectivity changes) in the post-blast network.

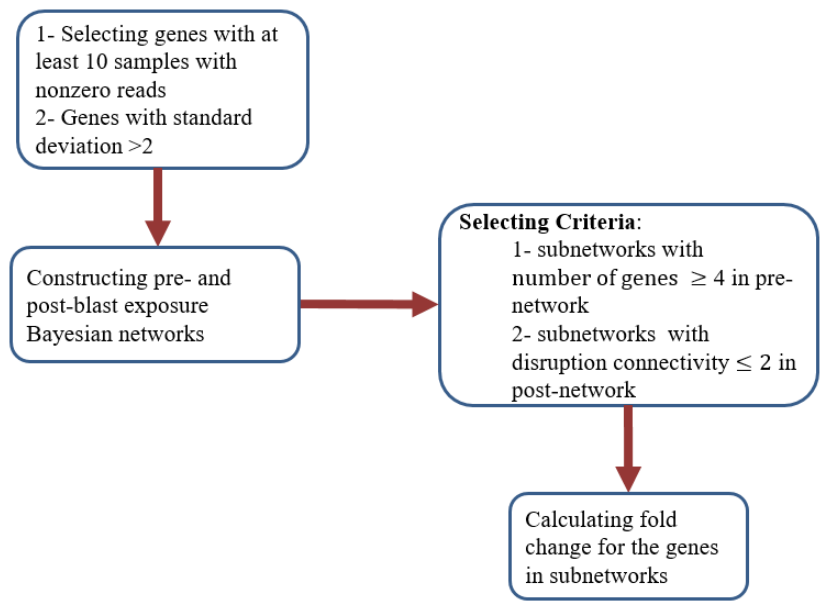

**Figure 2**- Strategy of selecting target subnetworks.

**Results**

To determine whether the effects of exposure waves produced by blast causes perturbations in gene-gene interactions, we compared the size (as defined by the number of genes within the subnetwork) and network topology in pre- and post-blast transcriptomic analyses (Figure 3). We observed the largest transcriptional perturbations in genes in subnetworks that contained 3, 4, and 5 genes, which showed a decrease in number of 18%, 27%, and 45%, respectively, following blast exposure.

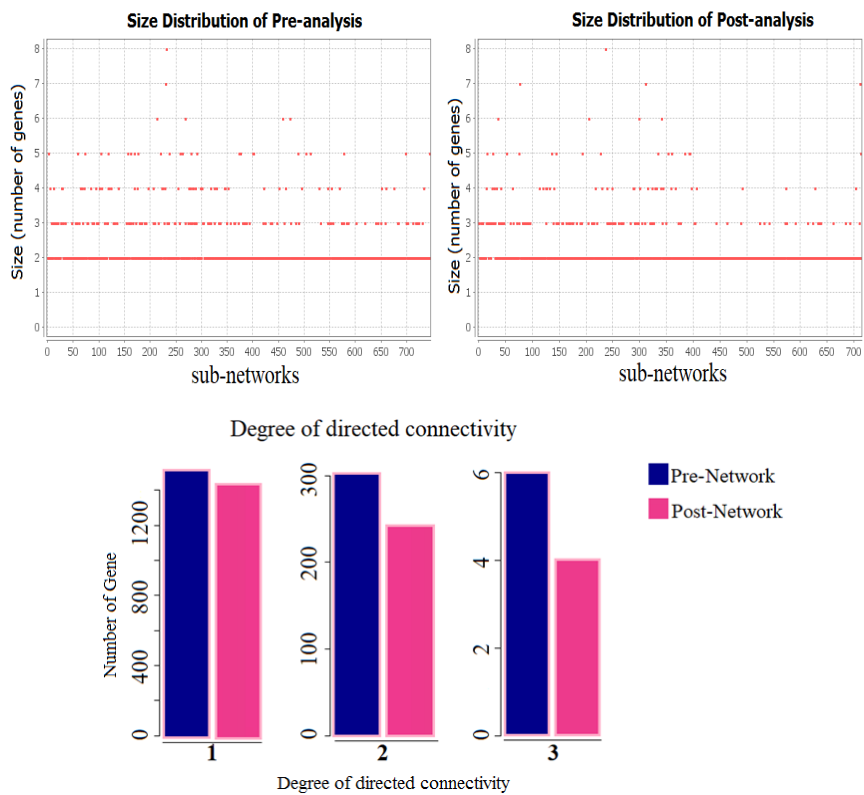

**Figure 3**- The influence of blast on transcriptional gene networks is measured by the degree of network connectivity and network size as defined by number of genes within a subnetwork. Top-left: Distribution of the size of subnetworks in pre-network analysis. Top-right: Distribution of the size of subnetworks in post-network analysis. Bottom-panel: Degree of directed connectivity in the pre- and post- data analyses.

In both networks, each gene interacted with at most 3 other genes directly. Overall, there was a reduction in direct gene-gene interaction in the post-blast exposure network presented in Figure 3, bottom-panel, as degree of directed connectivity.

Closer examination of these networks identified 4 biologically meaningful subnetworks based on the pipeline outlined in Figure 2. In order to identify those subnetworks that showed greatest transcriptional perturbations associated with blast, we prioritized those genes that showed highest fold change (FC) in gene expression pre- vs post-blast exposure. FC is computed as the average of the ratio of pre vs post transcriptomic levels of each sample and elucidates which gene in the network has major impact on dysregulation of these subnetworks. A FC > 1 indicates upregulation and a value of less than one indicates downregulation. Genes with relatively a large up/down regulation in the identified networks are visualized in Figure 4 (Supplementary 1 includes the FC of all genes in identified subnetwork).

Ingenuity pathway analysis (IPA, www.ingenuity.com) indicates that all genes in the identified subnetworks have common function related to organismal injury and abnormalities. Further details on the function and potential role in neurological disease of this interaction are presented in the discussion and supplementary 2.

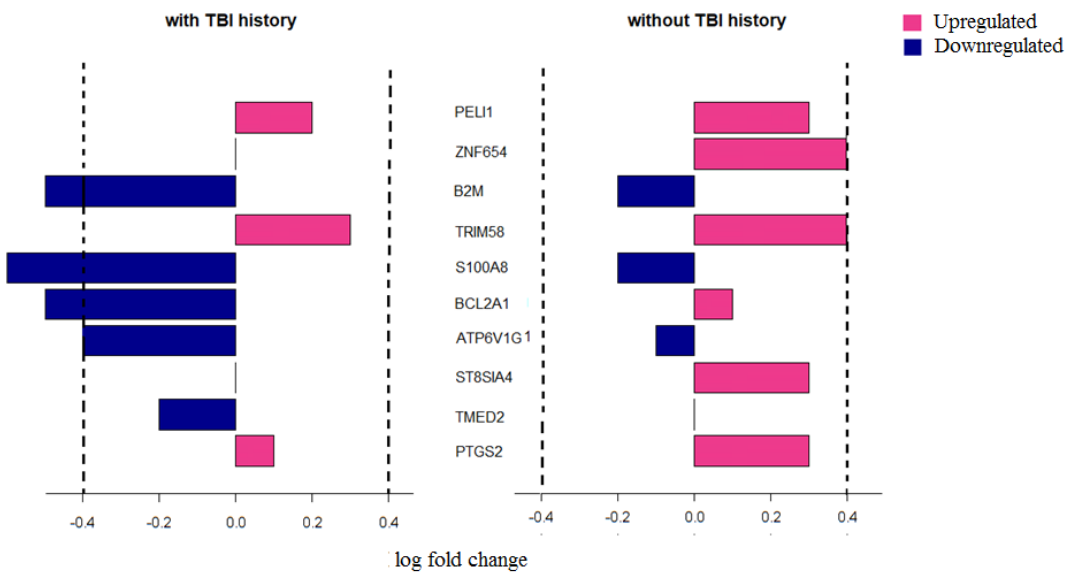

**Figure 4**- Genes showing greatest log FC in identified networks using the network selection strategy. Left bar plot depicts log fold changes in gene expression across individuals with lifetime history of TBI, whereas, right bar plot depicts log FC for corresponding genes in subjects without lifetime history of TBI. Vertical lines demarcating ±0.4 log FC correspond to thresholds of significance chosen.

In addition, we separated the samples into two subsets with and without self-reported lifetime history of TBI and calculated the fold change for each subset. Interestingly, upregulated genes due to blast exposure showed greater changes among individuals without TBI history. In contrast, downregulated genes showed greater changes among individuals with TBI history. This may be due to the differences between acute effect of blast exposure and accumulative effect of repeated blast exposure. We did not observe similar patterns of gene regulation in individuals with prior lifetime history of blast exposure (the breachers). This may be attributed to small sample sizes with only 8 individuals having no prior history of blast exposure in the entire data set.

**Characteristics of selected subnetworks**

**Subnetwork 1:** In the pre-blast network, subnetwork 1 consists of 4 genes; Pellino E3 Ubiquitin Protein Ligase 1 (*PELI1*), ST8 Alpha-N-Acetyl-Neuraminide Alpha-2,8-Sialyltransferase 4 (*ST8SIA4*), Integral Membrane Protein 2B (*ITM2B*), and Prostaglandin-Endoperoxide Synthase 2 (*PTGS2*). Among these genes (*PELI*1, *ITM2B* and *PTGS2*) are those which have been previously implicated in TBI (Huang et al., 2017; Wu et al., 2013; Graber, Costine, & Hickey, 2015). However, in the post-blast network, there are losses of connectivity between the genes (*ITM2B, PTGS2*) and (*ST8SIA4, ITM2B*) (Figure 5, top-left panel). Comparing FC expression levels with pre- vs post-blast networks shows that the blast has a greater effect on individuals without lifetime history of TBI. Figure 5, bottom-panel, shows the violin plot of gene transcripts with relatively higher fold change.

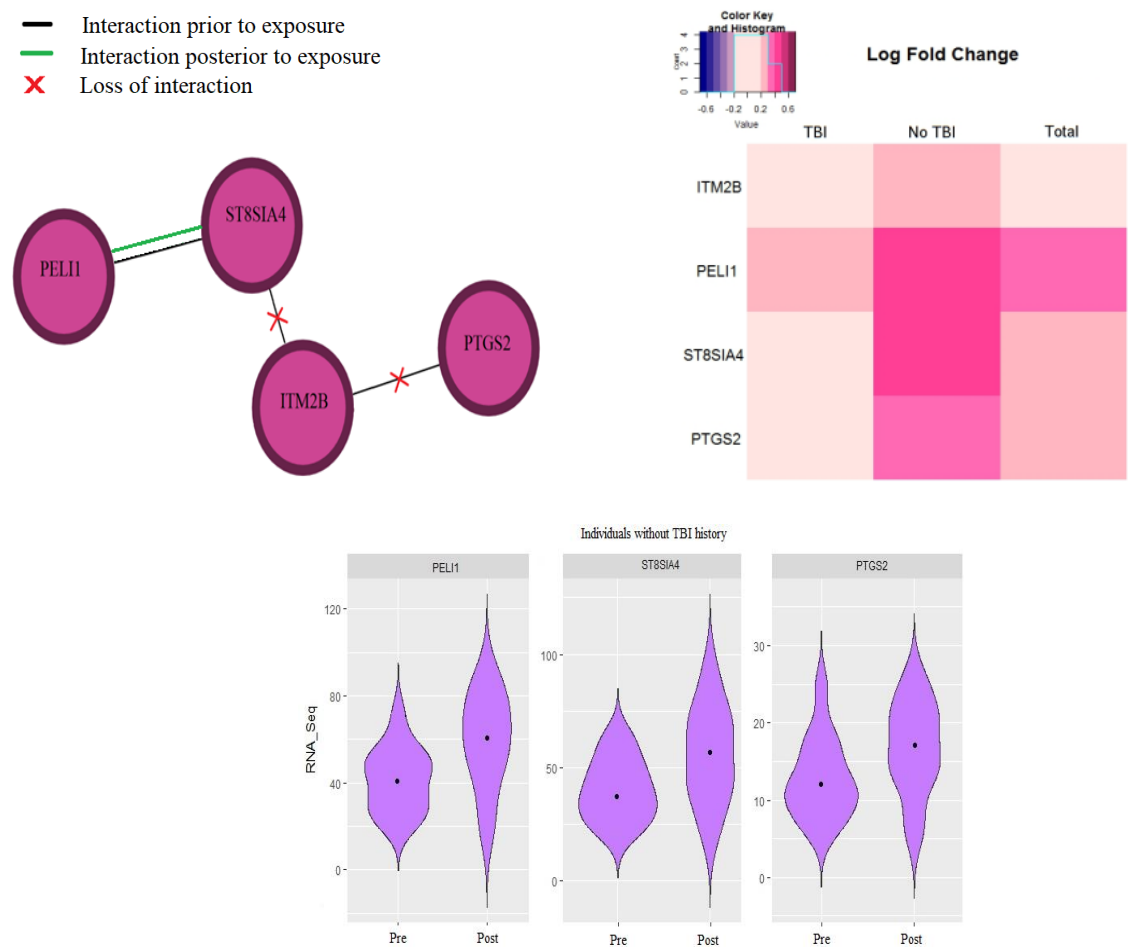

**Figure 5**- Top-left: Identified subnetworks; black/green links represent gene interactions prior/posterior to the training and red cross represents missing link in posteriori. Top-right: log FC of gene transcript levels for all individuals, with and without lifetime history of TBI. Bottom-panel: Violin plots of gene transcript levels from individuals without lifetime history of TBI for the genes with relatively higher fold change.

While we observed overexpression of *PELI1, ST8SIA4* and *PTGS2* in response to blast exposure in individuals without TBI history (Figure 5 bottom-panel), we conclude that induced transcript levels of these genes play a key role in disruption of interactions.

**Subnetwork 2:** Genes involved in this subnetwork are BCL2 Like 1 (*BCL2L1*), DDB1 and CUL4 Associated Factor 12 (*DCAF12*), Proteasome Inhibitor Subunit 1 (*PSMF1*) and Tripartite Motif Containing 58 (*TRIM58*). In the post-blast network, *PSMF1* did not show any interaction with other genes in the network (Figure 6, top -right). The analysis revealed that TBI associated gene *TRIM58* (Heinzelmann et al., 2014) has the highest FC in the subnetwork and interacts directly with *DCAF12* which shows slight changes in its expression level. Furthermore, overexpression of these genes influenced the relation between *PSMF1* and *DCAF12*. Figure 6, bottom-panel, shows the probability density of the data at different transcript values of *TRIM58* through violin plots.

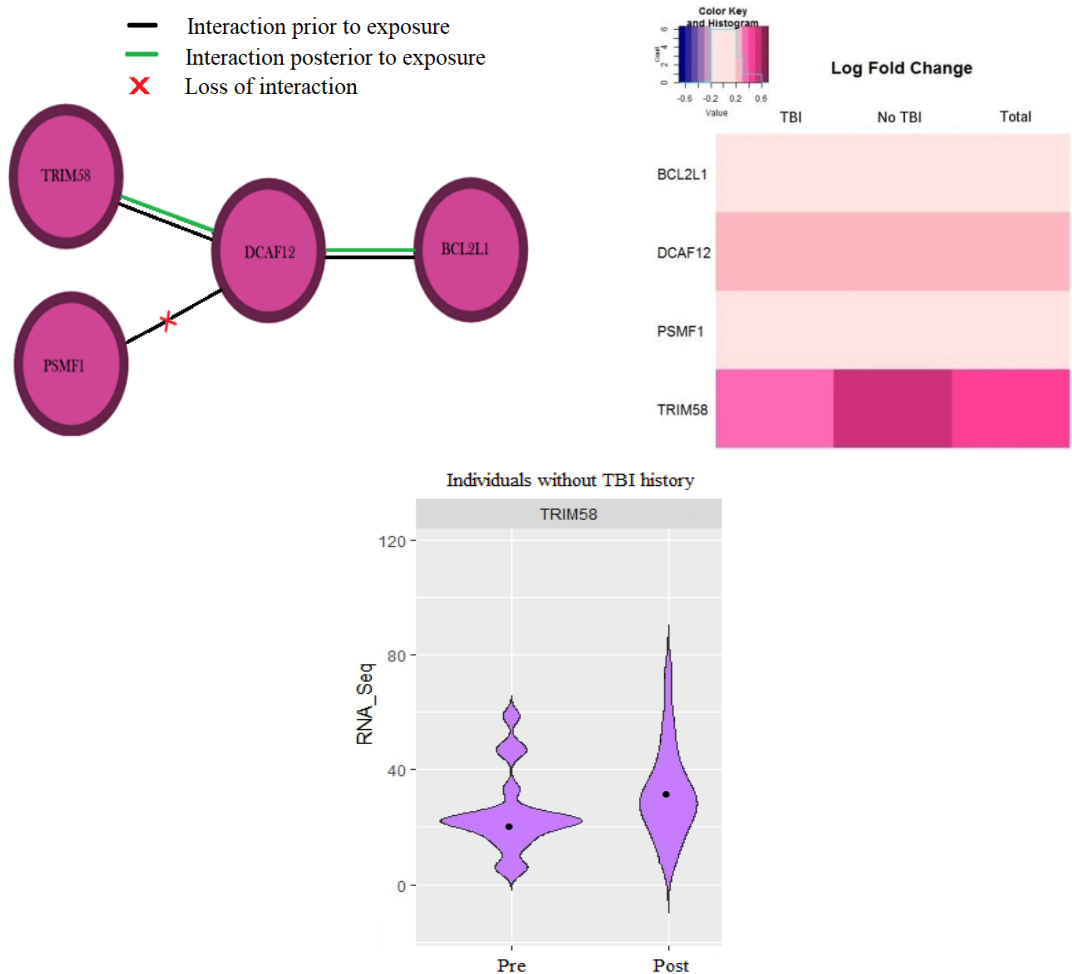

Figure 6- Top-left: Identified subnetworks; black/green links represent gene interactions prior/posterior to the training and red cross represents missing link in posteriori. Top-right: Log fold change of gene transcript levels for all individuals, with and without lifetime TBI history. Bottom-panel: Violin plots of gene transcript levels from individuals without lifetime TBI history for *TRIM58* with relatively higher fold change.

**Subnetwork 3**: This subnetwork comprises 5 genes, Beta-2-Microglobulin (*B2M*), ATPase H+ Transporting V1 Subunit G1 (*ATP6V1G1*), BCL2 related protein A1 (*BCL2A1*), lymphocyte antigen 96 (*LY96*), and S100 calcium binding protein A8 (*S100A8*). The effect of blast exposure has been represented in loss of connectivity in this subnetwork such that it is broken down into two smaller subnetworks (*B2M*, *ATP6V1G1*) and (*BCL2A1*, *LY96*, *S100A8*). All genes in this subnetwork are related to immune system and brain injury (see discussion). As shown in Figure 7, gene *BCL2A1* and *ATP6V1G1* stop interacting due to exposure effect. In addition, *S100A8* interacts indirectly with *LY96* despite its direct interaction in pre-blast network is lost upon exposure.

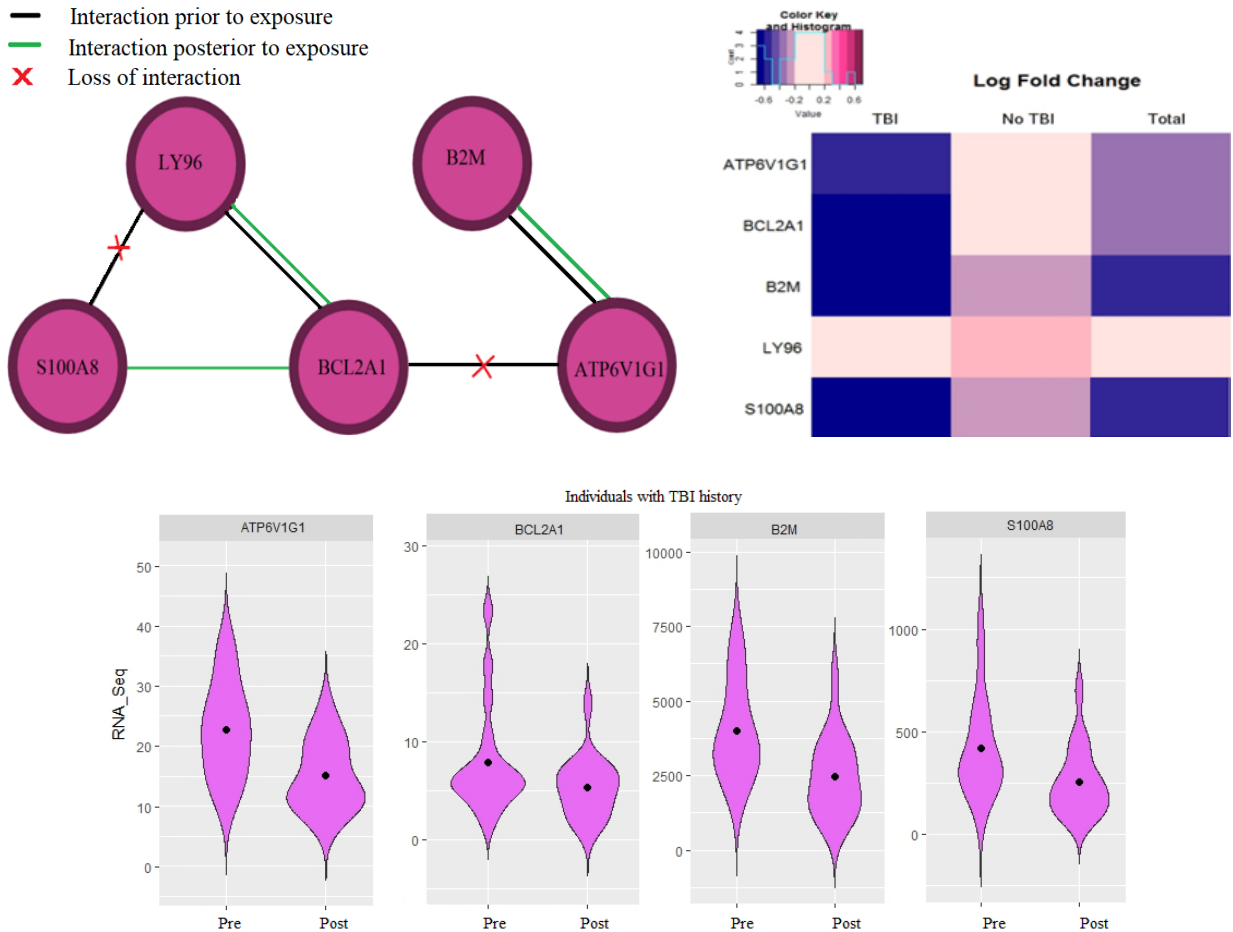

Figure 7- Top-left: Identified subnetworks; black/green links represent gene interactions prior/posterior to the training and red cross represents missing link in posteriori. Top-right: Log FC of gene transcript levels for all individuals, with and without lifetime TBI history. Bottom-panel: Violin plots of gene transcript levels from individuals with lifetime TBI history for genes with relatively higher fold change.

The alteration observed in this subnetwork downregulates gene expression after exposing to waves produced by blast. Figure 7, shows the effect of exposure on individuals with lifetime history of TBI is higher than individuals without lifetime TBI history. In addition, Figure 7, bottom-panel, displays probability densities of the genes with highest change through violin plots. This result may be an evidence of accumulated effect of repeated exposures to blast or traumatic conditions in biological reaction.

**Subnetwork 4:** It includes 5 genes; Peroxiredoxin 3 (*PRDX3*), Ring Finger Protein 139 (*RNF139*), Succinate Dehydrogenase Complex Subunit D (*SDHD*), Stress Associated Endoplasmic Reticulum Protein 1 (*SERP1*), and Transmembrane P24 Trafficking Protein 2 (*TMED2*). The post-blast transcriptomic analysis shows that *RNF139* interacts with Zinc Finger Protein 654 (*ZNF654*) while its interaction with *TMED2* and *SERP1* are disrupted in the network. The heatmap in Figure 8, top-right, displays upregulation of *ZNF654* in the group of individuals with no history of TBI while small FC can be seen for *TMED2* (downregulated). The violin plots in Figure 8 represents further details on changes of gene transcript levels.

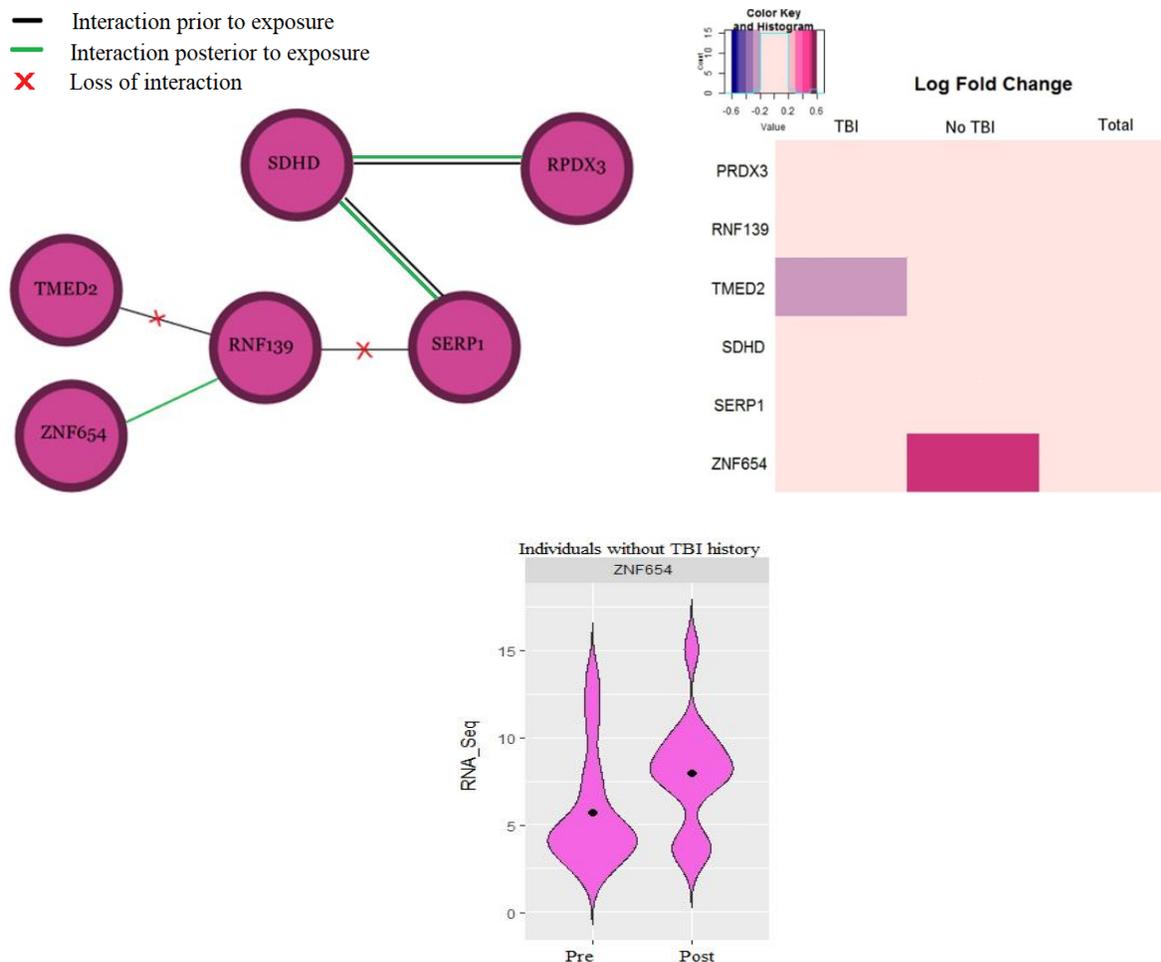

Figure 8- : Top-left: Identified subnetworks; black/green links represent gene interactions prior/posterior to the training and red cross represents missing link in posteriori. Top-right: Log FC of gene transcript

levels for all individuals, with and without lifetime TBI history. Bottom-panel: Violin plots of gene transcript levels from individuals without lifetime history of TBI for genes with relatively higher fold change.

**Discussion**

In this study, we investigated underlying relationships among transcript levels from RNA-seq data collected prior to and following the training program. Comparing the identified networks in each set of data provided insights into mechanisms of blast exposure impact on gene interactions and potentially TBI progress. This study generated blood-based biomarkers which are desirable with advantages of being cost- and time-effective while simultaneously being feasible at the population level compared with the collection of cerebrospinal fluid (CSF) or neuroimaging, (O'Bryant et al., 2016); (O'Bryant et al., 2017); (Lista, Faltraco, Prvulovic, & Hampel, 2013). Moreover, this study provided systematic knowledge and narrows down the search space for further studies regarding treatment and prevention. Therefore, these kinds of systems approaches are complementary to experimental studies especially in the age of big data.

One of the identified subnetworks includes 4 genes (*PELI1, ST8SIA4, ITM2B,* and *PTGS2*) that are overexpressed among individuals without lifetime TBI history. This subnetwork that is dysregulated immediately after two-week trainees can be a good target for preventing acute health effects of exposure to blast. In addition, all these genes are expressed in brain tissue and influence brain function through beta amyloid which is one of the causes of Alzheimer's disease (AD). While TBI can initiate long-term neurodegeneration processes leading to pathological features similar to AD (Ramos-Cejudo et al., 2018; Washington, Villapol, & Burns, 2016), this path could be seen as a link between TBI and AD.

Gene *PELI1* with the highest fold change in the subnetwork is known to play a pivotal role in inflammatory and autoimmune processes (Choe et al., 2016). Several possible phenotypes are observed depending on the extent of *PELI1* gene expression (Park J. et al. 2017). Interestingly, our analysis is consistent with their experimental observation since our study showed higher expression of *PELI1* on individuals without lifetime TBI history while *ST8SIA4* expression altered as its direct interacted gene.

*ST8SIA4* catalyzes the polycondensation of alpha-2,8-linked sialic acid required for the synthesis of polysialic acid. Molecular, cellular, and genetic studies implicate polysialic acid: in the control of cell-cell and cell-matrix interactions; intermolecular interactions at cell surfaces; and interactions with other molecules in the cellular environment (Hildebrandt H., & Dityatev A. 2013). Therefore, dysregulation in relation between *ST8SIA4* and *ITM2B* may be the cause of observing *ITM2B* immunoreactivity in extracellular deposits while glycosylation plays an important role in protein trafficking of *ITM2B*. The *ITM2B* deposition starts in early stages of the Alzheimer's disease (AD) pathology (Eliss et al., 1989) which supports our result since we showed that the genes in this subnetwork are overexpressed among individuals without TBI history.

*ITM2B* and *PTGS2* interaction is also disrupted while we observed overexpression of *PTGS2*. The gene *PTGS2* (*Cox-2*) is responsible for the prostaglandin biosynthesis involved in inflammation (https://www.genecards.org/cgi-bin/carddisp.pl?gene=PTGS2). Furthermore, *PTGS2* is the key enzyme that controls the production of Prostaglandin E (PGE), also involved in various inflammatory processes and degenerative disorders (Woodling et al. 2014). Overexpression of *PTGS2* elevates PGE2 levels. As suggested by Ellis et al. (1989), increased levels of PGE2 after injury may be a body-wide response including brain. PGE2 in brain does dependently inhibit macrophage-mediated phagocytosis of aggregated Aβ (Fox et al., 2015; Aronoff, Canetti, & Peters-Golden, 2004) and elevated levels of PGE2 have been observed in brains of patients with AD. Several lines of evidence indicate that PGE2 inhibits the microglial-mediated phagocytosis of aggregated Aβ through the prostaglandin $E_2$ receptor subtype 2 (EP2) receptor. Microglia lacking EP2 also demonstrated enhanced phagocytosis of Aβ plaques presented on hippocampal sections from patients who died with Alzheimer's disease (Fox et al., 2015).

*TRIM58* is one of the identified genes in subnetwork 2 that previously has shown differential expression in case control study of TBI (Heinzelmann et al., 2014). Heinzelmann et al. found reduction of *TRIM58* in individuals with blast-TBI. However, we did not observe significant changes of *TRIM58* among individuals with lifetime history of TBI. Instead, we observed upregulation of *TRIM58* among people without lifetime history of TBI.

The other subnetwork with 5 genes (*B2M*, *ATP6V1G1*, *BCL2A1*, *LY96*, and *S100A8*) is related to the immune system. Except *LY96*, all other genes are low expressed among individuals with lifetime history of TBI. Thus, we can hypothesize that this path represents the accumulated effect of repeated exposures to blast or traumatic conditions on immune system.

Genes *B2M* and *ATP6V1G1*, which interact prior to and following the training program, are related to the innate immune system. Exposing to the blast wave resulted in a low expression level of both genes such that low expression of *B2M* causes low level of *MHC* class I antigens which influence a proper response of immune system (Safa, Saadatzadeh, Gadol, Pollok, & Vishehsaraei, 2016). Low expression of *ATP6V1G1* induces cell death since *ATP6V1G1* encodes a component of vacuolar ATPase, a multi-subunit enzyme necessary for intracellular processes as protein sorting, zymogen activation, receptor-mediated endocytosis, and synaptic vesicle proton gradient generation (Safa et al., 2016).

Three other genes in this subnetwork (*BCL2A1*, *LY96*, *S100A8*) continued interacting after the training program although *S100A8* interacts *with LY96* indirectly through *BCL2A1*. *S100A8* and *BCL2A1* with relatively large downregulation after exposing to blast, are involved in the immune system processes. Expression of *BCL2A1* with important role in immune system protects cells from various death stimuli. Down regulation of *BCL2A1* causes loss of survival of pre–T cell (Mandal et al., 2005). *S100A8* also stimulates innate immune cells and acts as an alarm in or a danger associated molecular pattern (DAMP) molecule. It is involved in the regulation of a number of cellular processes such as cell cycle progression and differentiation. Interestingly, protein of *S100A8* interacts with *LY96* (protein *MD-2*) with a role in the innate immune system (Deguchi et al., 2016) as we observed in identified network in this study. This interaction breaks after the training program as the effect of exposing to the blast waves. Therefore, this study suggests that preventing dysregulation of identified interaction may reduce the effect of exposing to traumatic conditions on immune system.

In this research, we observed that the effect of traumatic conditions on people varies depending on their life time history of TBI. Although, limited sample prevents us to have group specific analysis for individuals with and without lifetime TBI history. Group specific analyses will help to find specific pathways for each group separately, and consequently have better prevention approaches. In addition, increasing the number of samples, leads to identify modules and longer pathways through network analysis.

## Acknowledgements

The author would like to thank Dr. Fatemeh Haghighi for providing the data and comments on the manuscript.

**Supplementary 1:**

Table 1: Average of log10 FC for individuals with/out TBI history and the total in subnetwork 1

|  | *ITM2B* | *PELI1* | *ST8SIA4* | *PTGS2* |
|---|---|---|---|---|
| **With TBI history** | 0.04 | 0.16 | 0.04 | 0.05 |
| **Without TBI history** | 0.20 | 0.34 | 0.30 | 0.27 |
| **Total** | **0.11** | **0.24** | **0.15** | **0.14** |

Table 2: Average of log10 FC for individuals with/out TBI history and the total in subnetwork 2

|  | *BCL2L1* | *DCAF12* | *PSMF1* | *TRIM58* |
|---|---|---|---|---|
| **With TBI history** | 0.06 | 0.19 | 0.12 | 0.29 |
| **Without TBI history** | 0.06 | 0.20 | 0.07 | 0.37 |
| **Total** | **0.06** | **0.20** | **0.10** | **0.32** |

Table 3: Average of log10 FC for individuals with/out TBI history and the total in subnetwork 3

|  | *ATP6V1G1* | *BCL2A1* | *B2M* | *LY96* | *S100A8* |
|---|---|---|---|---|---|
| **With TBI history** | -0.42 | -0.47 | -0.54 | -0.13 | -0.59 |
| **Without TBI history** | -0.09 | 0.06 | -0.19 | 0.17 | -0.15 |
| **Total** | **-0.28** | **-0.25** | **-0.39** | **-0.01** | **-0.41** |

Table 4: Average of log10 FC for individuals with/out TBI history and the total in subnetwork 4

|  | *PRDX3* | *RNF139* | *TMED2* | *SDHD* | *SERP1* | *ZNF654* |
|---|---|---|---|---|---|---|
| **With TBI history** | -0.12 | -0.13 | -0.21 | -0.10 | -0.11 | -0.04 |
| **Without TBI history** | 0.07 | 0.14 | -0.04 | 0.10 | 0.14 | 0.37 |
| **Total** | **-0.04** | **-0.02** | **-0.14** | **-0.02** | **-0.01** | **0.13** |

**Supplementary 2:**

IPA network for genes in subnetwork 1:

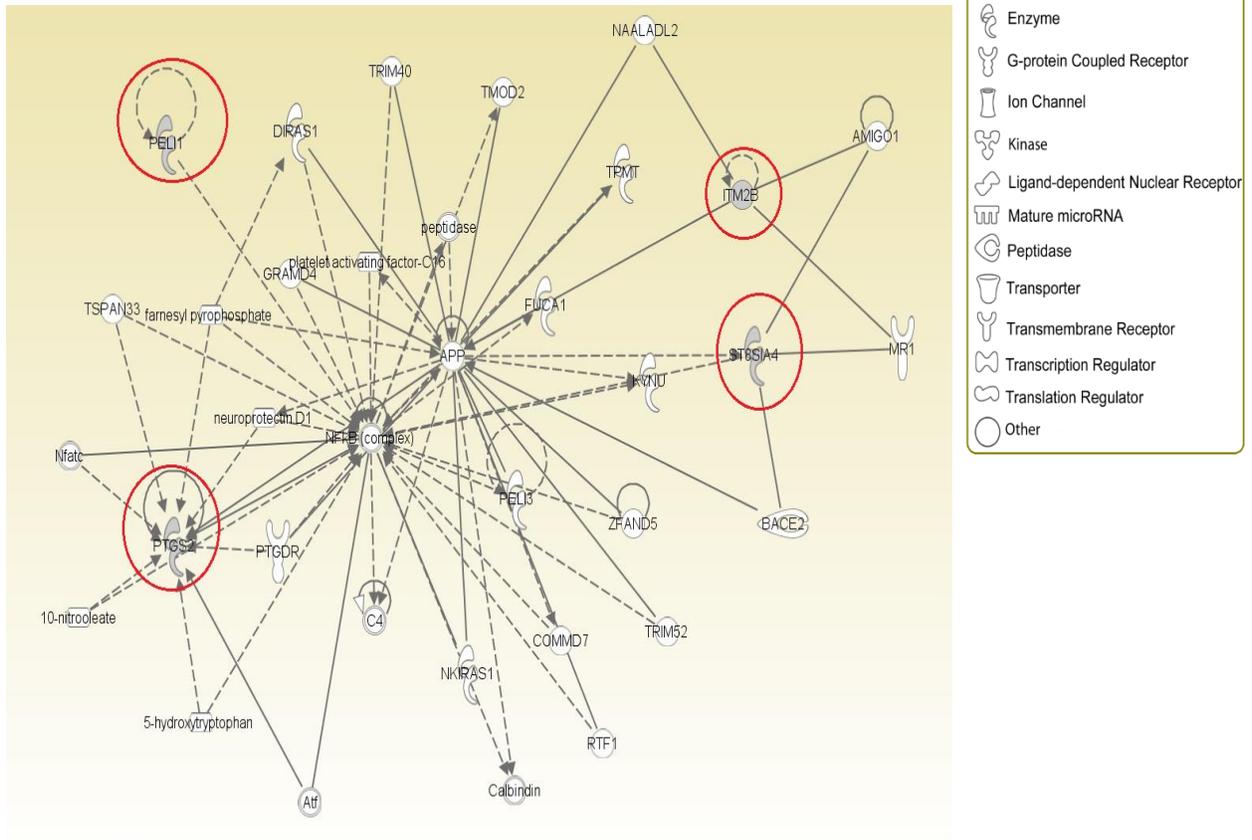

IPA network for genes in subnetwork 2:

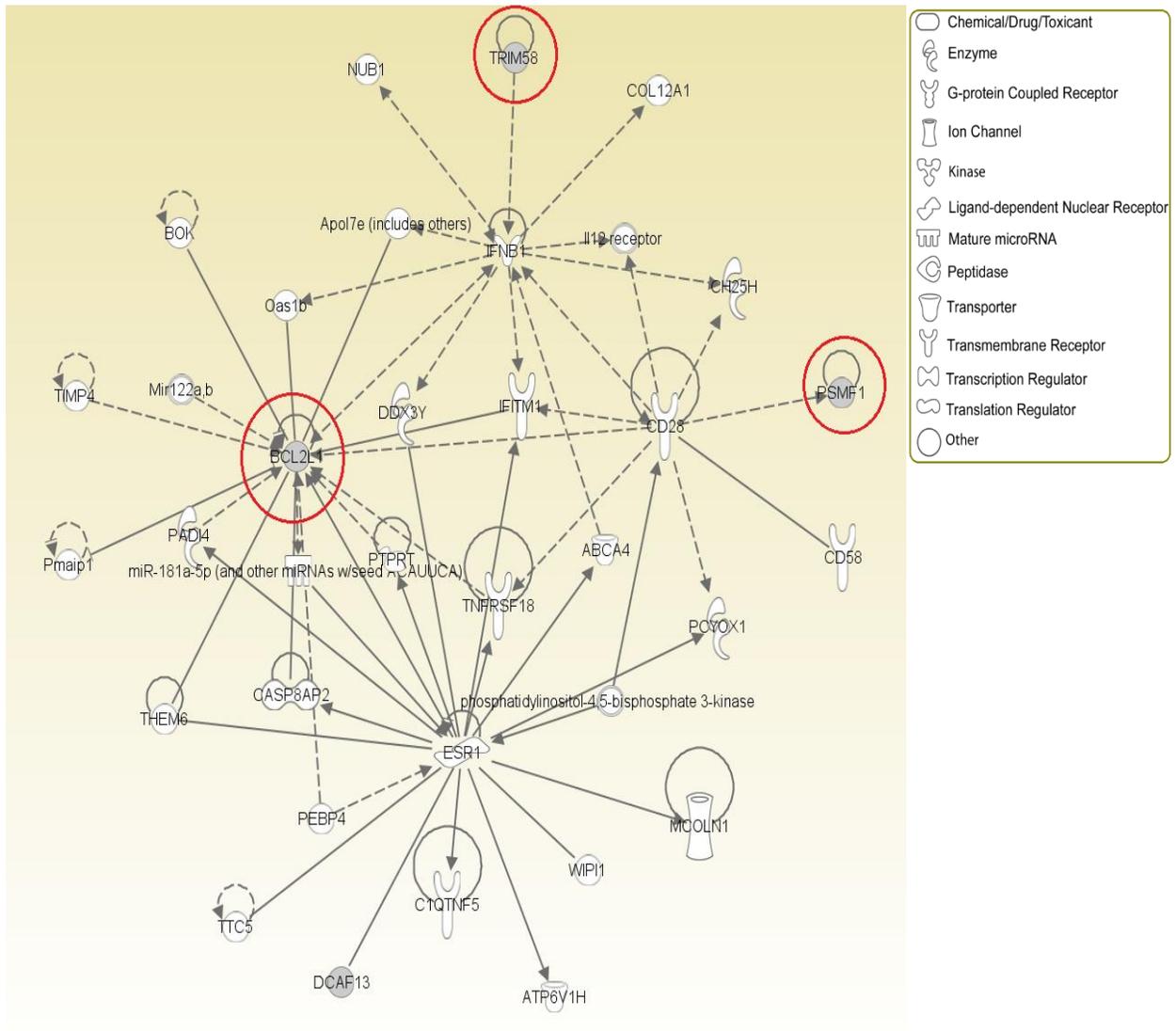

IPA network for genes in subnetwork 3:

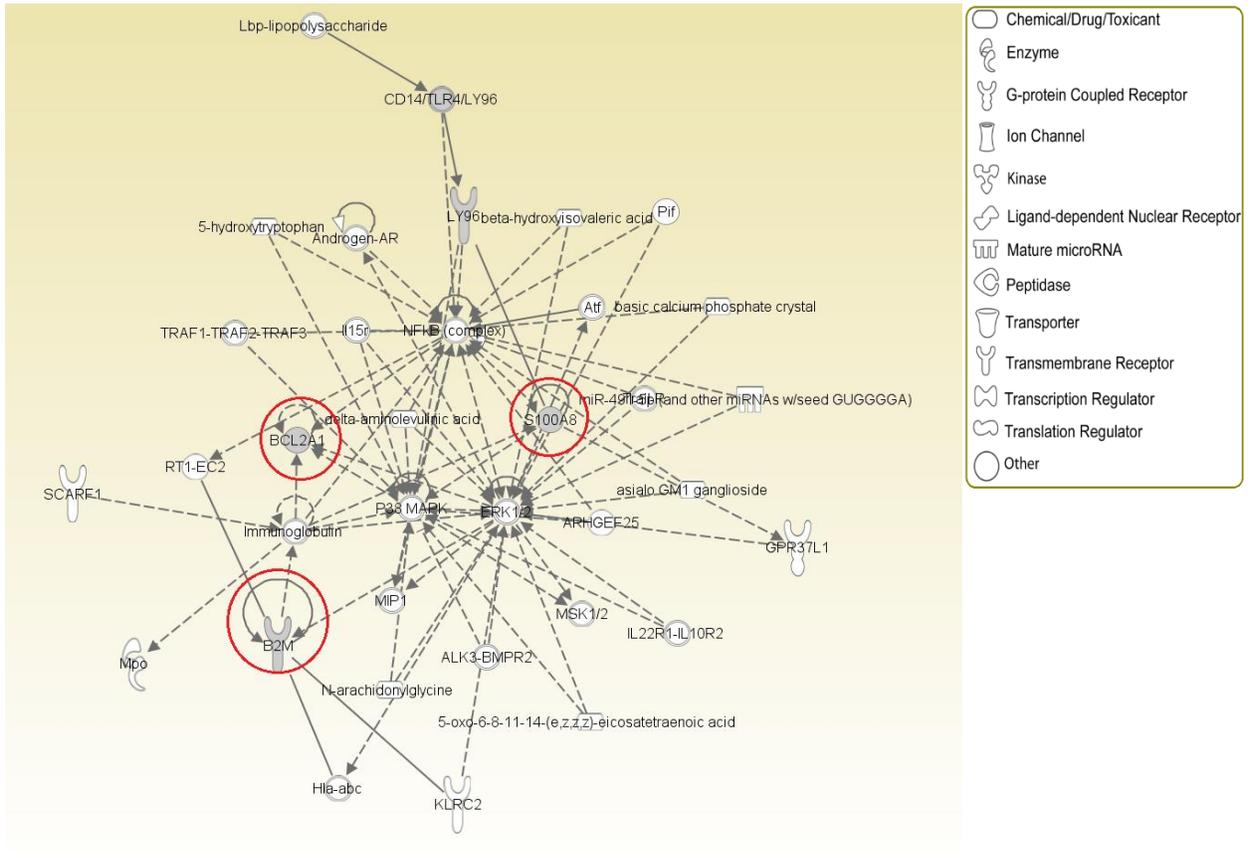

IPA network for genes in subnetwork 4:

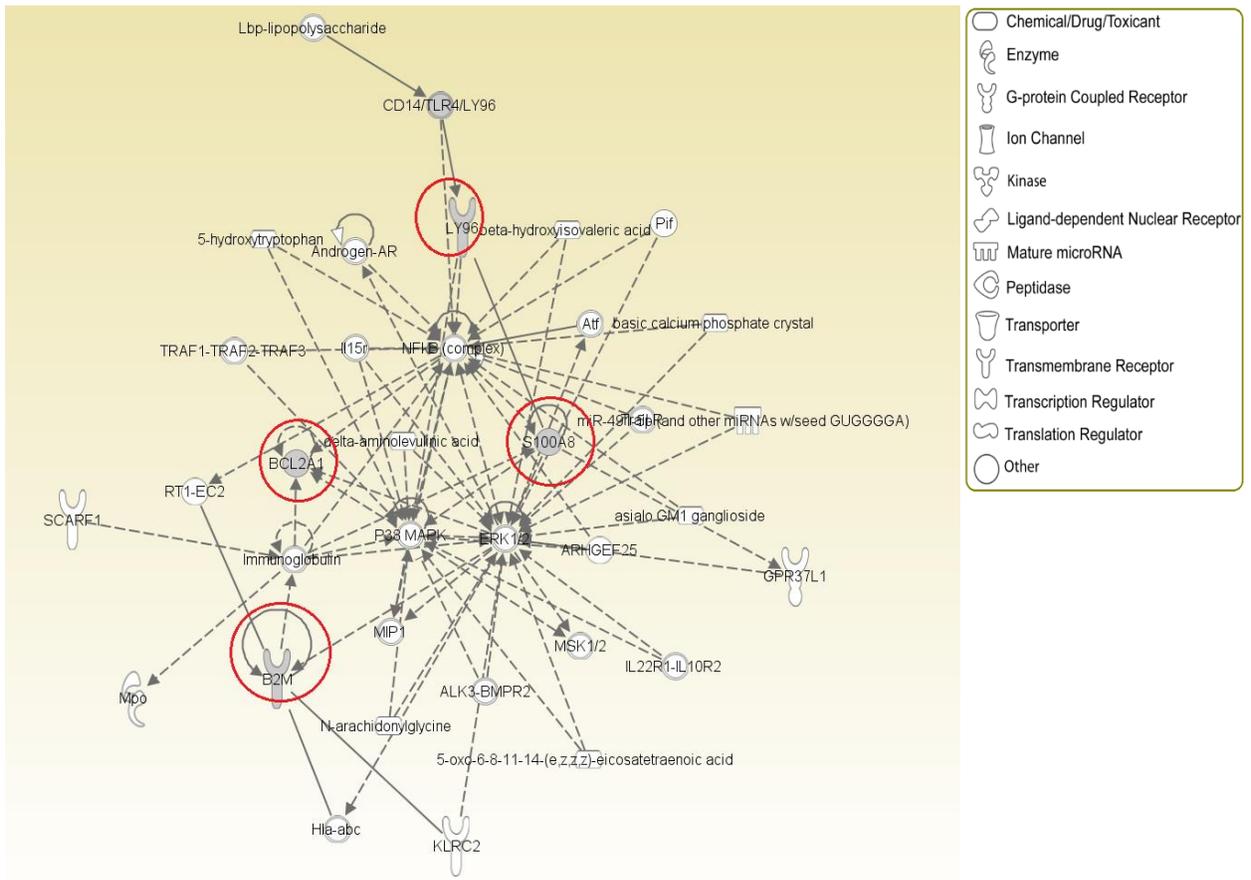